\documentclass[twocolumn,preprintnumbers,amsmath,amssymb,floatfix,nofootinbib,superscriptaddress,nobibnotes,showkeys]{revtex4-1}
\usepackage{amsmath,amssymb,latexsym,amsthm,ulem}
\usepackage{natbib} % reference manager
\usepackage[breaklinks=true]{hyperref}
\usepackage{breakcites}
\usepackage{cleveref}
\usepackage{verbatim}
\usepackage{wrapfig}
\usepackage{times}
\usepackage{epsfig}
\usepackage{graphicx}
\usepackage{color}
\usepackage{dcolumn}
\usepackage{bm}
\usepackage{setspace}
\usepackage{epstopdf}
\usepackage[loose,nice]{units} 
\usepackage{mwe} 
\usepackage{subfloat}
\usepackage{subcaption}
\usepackage{ragged2e}
\DeclareCaptionJustification{justified}{\justifying}
\usepackage{MnSymbol}
\definecolor{limegreen}{rgb}{0.2,0.8,0.2}
\definecolor{orange}{rgb}{1.0, 0.55, 0.0}
\usepackage{float}	
\usepackage{lipsum}
\usepackage{footnote}
\usepackage{titlesec}
\usepackage{tikz}
\usetikzlibrary{shapes}
\usetikzlibrary{plotmarks}
\usepackage[none]{hyphenat}

\newcommand{\markerfour}{\raisebox{0.5pt}{\tikz{\node[draw,scale=0.5,regular polygon, regular polygon sides=4,fill=none](){};}}}
\newcommand{\markerfive}{\raisebox{0.5pt}{\tikz{\node[draw,scale=0.5,diamond,fill=none](){};}}}
\newcommand{\markersix}{\raisebox{0.5pt}{\tikz{\node[draw,scale=0.6,circle,fill=none](){};}}}
\newcommand\solidrule[1][0.5cm]{\rule[0.5ex]{#1}{.4pt}}
	
	\newcommand\dashedrule{\mbox{%
	\solidrule[2mm]\hspace{2mm}\solidrule[2mm]\hspace{2mm}\solidrule[2mm]}}
	
	\titlespacing\paragraph{0pt}{12pt plus 4pt minus 2pt}{0pt plus 2pt minus 2pt}
\newcommand \E[1] {Eq.~(\ref{#1})}

\newcommand{\RNum}[1]{\uppercase\expandafter{\romannumeral #1\relax}}

\begin{document}

\title{Solidification Characteristics of Laser-Powder Bed Fused AlSi10Mg: Role of Building Direction}

%%%
\author{Hossein Azizi}
\email{haziz@unb.ca}
\affiliation{Marine Additive Manufacturing Centre of Excellence (MAMCE), University of New Brunswick, Fredericton, NB, Canada}
\affiliation{Department of Physics, Centre for the Physics of Materials, McGill University, Montreal, QC, Canada.}
\author{Alireza  Ebrahimi}
\affiliation{Marine Additive Manufacturing Centre of Excellence (MAMCE), University of New Brunswick, Fredericton, NB, Canada}%
\author{Nana Ofori-Opoku}
\affiliation{Computational Techniques Branch, Canadian Nuclear Laboratories, Chalk River, ON, Canada}%
\author{Michael Greenwood}
\affiliation{CanmetMATERIALS, Natural Resources Canada, 183 Longwood Road south, Hamilton, ON, Canada }%   
\author{Nikolas Provatas}
\affiliation{Department of Physics, Centre for the Physics of Materials, McGill University, Montreal, QC, Canada.}%
\author{Mohsen Mohammadi}
\affiliation{Marine Additive Manufacturing Centre of Excellence (MAMCE), University of New Brunswick, Fredericton, NB, Canada}%
%%%

%%%
\begin{abstract}
In this work, the effect of building direction on the microstructure evolution of laser-powder bed fusion (LPBF) processed AlSi10Mg alloy was investigated.  The building direction, as shown in experimentally fabricated parts, can influence the solidification behavior and promote morphological transitions in cellular dendritic microstructures. 
We develop a thermal model to systemically address the impact of laser processing conditions, and building direction on the thermal characteristics of the molten pool during laser processing of AlSi10Mg alloy. We then employ a multi-order parameter phase field model to study the microstructure evolution of  LPBF-AlSi10Mg in the dilute limit, using the underlying thermal conditions for horizontal and vertical building directions as input. The  phase field model employed here is designed to simulate solidification using heterogeneous nucleation from inoculant particles allowing to take into account morphological phenomena including the columnar-to-equiaxed transition (CET). The phase field model is first validated against the predictions of the previously developed steady-state CET theory of Hunt \cite{hunt1984steady}. It is then used under transient conditions to study microstructure evolution, revealing that the nucleation rate is noticeably higher in the horizontally built samples due to larger constitutional undercooling, which is consistent with experimental observations. 
We further quantify the effect of building direction on the local cooling conditions, and consequently on the grain morphology.
\end{abstract}
%%%


\maketitle

\section{Introduction}
\label{sec:1}
Additive manufacturing (AM), or 3D printing, is a material fabrication process whereby material feed stock in the form of powder or wire is gradually added in a layer-by-layer fashion. The added feed stock is melted selectively by a focused heat source, which then solidifies in a subsequent heating-cooling process until a resultant part is complete \cite{debroy2017additive, herzog2016additive}.   Among the available AM processes, laser-powder bed fusion (LPBF) has been widely accepted as a new paradigm for the design and production of high performance complex components. 
 This is due to its unique features, such as fast solidification rate, short manufacturing times, and controlled melting and solidification processes \cite{li2017selective,asgari2017microstructure}. This process is widely known to be well-suited for processing aluminum alloys, in particular AlSi10Mg \cite{olakanmi2015review}. LPBF-AlSi10Mg has attracted much attention recently due to its mechanical and structural properties for applications in aerospace, automotive, and marine industries \cite{hadadzadeh2018columnar,hadadzadeh2019role}. 

Significant effort has been made to study the microstructure evolution during AM processes \cite{wang2015grain}.  As evidenced by many studies, despite the complexities involved in AM processes, the evolution of grain structures for a given set of conditions can be largely controlled by an effective thermal gradient $G$, solidification rate $V$, and undercooling $\Delta T$ \cite{kurz1986theory}. Two widely observed solidification microstructures in AM processes are columnar and equiaxed dendritic structures \cite{wang2015grain}.  

Under certain conditions equiaxed dendrites nucleate in the constitutionally undercooled liquid adjacent to the solidification front. If the number of nucleated grains is sufficiently large columnar grains terminate with the formation of equiaxed grains and a columnar-to-equiaxed transition (CET) takes place \cite{gaumann2001single}.   There is some evidence that equiaxed grains may form due to homogeneous nucleation in the supercooled liquid \cite{parimi2014microstructural} or heterogeneously on partially melted powders or added refractory particles, i.e. inoculants \cite{antonysamy2013effect,wang2015grain,wang2011morphology}. In this study, we solely consider heterogeneous nucleation as the dominant mechanism in the formation of spontaneous nuclei.

\begin{figure}[!h]
\centering
\includegraphics[scale=.7]{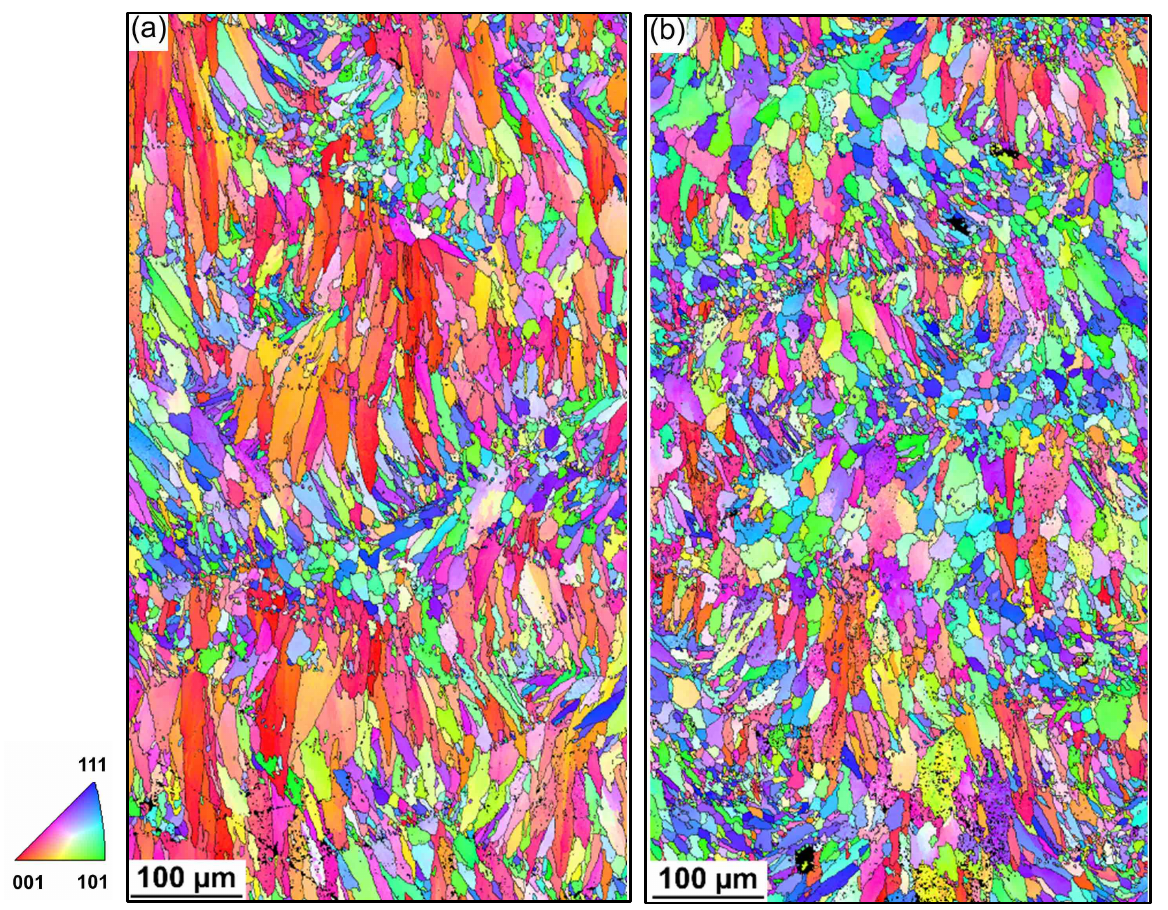}
\caption{EBSD IPF-Z images of samples with (a) vertical build and (b) horizontal build directions superimposed on corresponding grain boundaries. The figure is adopted from \cite{hadadzadeh2018columnar}}
\label{fig:1}  
\end{figure}

In recent years, numerous experimental and computational studies have been carried out to investigate alloy grain morphologies resulted from the LPBF process \cite{thijs2013fine,zhou2015textures,rao2016influence,li2017selective,hadadzadeh2018columnar}. For instance, Hadadzadeh et al. \cite{hadadzadeh2018columnar} conducted an experimental study to inspect the microstructure of AlSi10Mg alloy processed by LPBF. Their observations clearly confirmed that changes to the build direction can affect both columnar to equiaxed ratio and the texture of an LPBF-AlSi10Mg alloy. These effects are clearly illustrated in typical electron backscatter diffraction (EBSD) images for horizontal and vertical samples, an example of which is shown in Fig.~\ref{fig:1}.

In this work we make the following approximations to isolate our study on the salient properties of  LPBF-AlSi10Mg solidification:
\begin{enumerate}
    \item We focus on the solidification of a binary Al-Si alloy. The concentration of magnesium in AlSi10Mg is low  ($0.33 \,{\rm wt}.\%$).  	Moreover, the local maximum temperature of the laser $T^{\rm max}_{\rm L}$ ($\sim$1900 K) exceeds the boiling 	
    	temperature of magnesium $T^{\rm boil}_{\rm Mg}$ (1363 K), which further reduces the concentration of magnesium. 
	Therefore, the effect of magnesium on the microstructure evolution appears to be minimal and is therefore omitted from 
	the simulations.
    \item We study the solidification of binary Al-Si alloy in the dilute limit. The solute (Si) concentration is set to a very small value ($0.5 \,{\rm wt}.\%$) which is more consistent with the dilute alloy limit assumptions. 
\end{enumerate}
We will show that our multi-order parameter phase field model coupled with transient thermal conditions relevant to the LPBF process is sufficient to explain the qualitative mechanisms that building direction affects resulting microstructure.
%%%
\section{Phase field model }
\label{sec:2}
We employ an adaptation of the model of Ofori-Opoku et al. \cite{ofori2010quantitative} formulated from a grand potential functional~\cite{plapp2011unified}, which allows us to model the evolution of the chemical potential field alongside the order parameters representing grains. The details of the model are found in \cite{greenwood2018quantitative,K_shampur}.
In this model the order parameter vector is defined as $\vec{\phi}(\vec {r}) = \big(\phi_{1}(\vec{r}),\phi_{2}(\vec{r}),...,\phi_{N}(\vec{r})\big)$, whose components are bounded by 0 and 1, representing the bulk of liquid and solid phases, respectively. Each order parameter $\phi_{i}$ represents one of $N$ distinct solid grain orientations or crystal phases at a particular point in a volume. Through their interactions, they always satisfy $\phi_{1}+\phi_{2}+...+\phi_{N} \leq 1$ \cite{greenwood2018quantitative}. The solid-liquid interface is represented as a region with finite (but non-zero) thickness $W_{i}$ in which the order parameter $\phi_{i}$ varies, continuously, between its two bulk values. 
The grand potential functional governing the properties of a multi-grain dilute binary system can be written in the following form \cite{plapp2011unified}:
\begin{multline}
\label{eq:freeEnergyFunctional}
 \Omega[\vec{\phi},\mu] = \int {dv}\,\biggl\{\omega_{\rm int}(\vec{\phi},\vec{\nabla}{ \phi}) + \\
+ \sum_{\alpha=1}^{N}g_{\alpha}(\vec{\phi})\,\omega^{\alpha}({\mu}) +  \biggl[1-\sum_{\alpha=1}^{N}g_{\alpha}(\vec{\phi})\biggl]\,\omega^{l}({\mu})\biggl\},
\end{multline}
where $\omega$ and $\mu$ are the grand potential density and chemical potential, respectively, $l$ is the liquid phase and index $\alpha$ runs over solid phases or orientations.  The first term in Eq.~(\ref{eq:freeEnergyFunctional}) accounts for the interaction energy between order parameters which can be written in the following form, 
\begin{multline}
\omega_{\rm int}(\vec{\phi},\vec{\nabla}{ \phi}) = \sum_{\alpha=1}^{N}\frac{\sigma^{2}_{\alpha}}{2}{\big|\nabla{\phi_{\alpha}}\big|}^{2}\\
+\sum_{\alpha=1}H_{\alpha}f_{\rm {DW}}(\phi_{\alpha})+\sum_{\alpha, \alpha\neq \beta}\omega_{\alpha\beta}\Psi(\phi_{\alpha},\phi_{\beta}),
\end{multline}
where $\sigma_{\alpha}$ are constants that set the solid-liquid interface energies for each phase $\alpha$, $H_{\alpha}$ defines the nucleation energy between solid $\alpha$ and liquid, $f_{\rm DW}$ is a double-well potential (with minima at $\phi_{\alpha}$ = 0,1), and the term $\Psi(\phi_{\alpha},\phi_{\beta})\propto{\phi_{\alpha}}^{2}{\phi_{\beta}}^{2}+\dots$ contains polynomial order interaction terms between different order parameters. 
In this study, we retain only the second order pair-wise term as we are mainly interested in morphological patterning of LPBF microstructure as a result of nucleation and free-growth kinetics of competing grains during solidification. Other forms of this interaction term have been examined in \cite{K_shampur}. The remaining terms in Eq.~(\ref{eq:freeEnergyFunctional}) contain the grand-potential densities of the bulk solid ($\omega^\alpha$) and liquid ($\omega^l$) phases, where the functions $g_{\alpha}(\vec{\phi})$ interpolate the local grand potential density between phases via the order parameter components $\phi_{\alpha}$.

The dynamics of solidification in the grand potential formalism is described by the coupled evolution of all order parameters and the chemical potential. The dynamics of each order parameter $\phi_{\alpha}$ is given by 
\begin{equation}
\label{eq:evolutionPhi}
\frac{\partial{\phi_{\alpha}}}{\partial{t}} = -M_{\phi_{\alpha}}\,\frac{\delta \Omega}{\delta \phi_{\alpha}} + \xi_{\phi},
\end{equation}
and chemical potential of the solute species (relative to the solvent), $\mu$, is given by
\begin{multline}
\label{eq:evolutionmu}
\frac{\partial{\mu}}{\partial{t}} = \frac{1}{\chi}\biggl[\nabla\cdot\,\bigg(M(\vec{\phi},c)\,\nabla\mu\bigg)-\\
\sum_{\alpha}g^{'}_{\alpha}(\vec{\phi})\,\big(c^{\alpha}(\mu)-c^{l}(\mu)\big)\,\frac{\partial{\phi_{\alpha}}}{\partial{t}} - \nabla\cdot\vec{\zeta}\biggl]
\end{multline}
Here, $M_{\phi_{\alpha}}$ sets a suitable time scale for $\phi_{\alpha}$, $M(\vec{\phi},c)$ is Osanger-type mobility coefficient for mass transport, and $\chi$ is the susceptibility parameter that is defined as $\partial{c}/\partial{\mu}$. 
In practice, a non-variational term is added in Eq.~(\ref{eq:evolutionmu}), to correct for anomalous solute-trapping effect arising from the use of an interface width in the model that is much larger than the actual physical value \cite{plapp2011unified}. The stochastic fields $\xi$ and $\vec{\zeta}$ in Eqs.~(\ref{eq:evolutionPhi})-(\ref{eq:evolutionmu}) respectively account for thermal fluctuations in order parameters and noise flux governing fluctuations in solute concentration, and satisfy \textit{ fluctuation dissipation theorem}~\cite{chaikin2000principles}.

%%%
\section{Nucleation}
\label{sec:3}
In this study, we consider heterogeneous nucleation on partially melted powders (inoculants). This, as evidenced in experimental studies, is typically accepted as the dominant mechanism of heterogeneous nucleation \cite{xi2019effect}. In order to examine spontaneous nucleation in an undercooled liquid, we adopt a similar approach as discussed in \cite{castro2003phase}. In this method, the energy barrier of solid-liquid nucleation within a volume $\Delta V$ (surface $\Delta A$ in 2D) of an inoculant, is set to a value corresponding to heterogeneous nucleation for a given contact angle $\theta_c$. The free energy barrier for heterogeneous nucleation for a given set of conditions (solute concentration and temperature) is then calculated according to classical nucleation theory,
\begin{equation}
\label{eq:freeEnergyBarrier}
\Delta F^{*}=\frac{16\,\pi}{3}\,\frac{h(\theta_c)\,\gamma^3}{{\Delta f_{\nu}} ^2},
\end{equation}
 where $\gamma$ is the interfacial energy ($\rm J/m^{2}$) , $h$ is a heterogeneous pre-factor that depends on the contact angle $\theta_c$, and $\Delta f_{\nu}$ is the bulk free energy change per unit volume ($\rm J/m^{3}$).  The contact angle $\theta_c$ depends on the interfacial energy between the liquid and the surface of an inoculant particle. In our simulations, the contact angles of inoculants are assigned randomly to reflect the random nature of inoculants.
\subsection{Incorporating thermal fluctuations in the phase field model}
Here, we consider nucleation arising from thermal fluctuations in the order parameter equations, Eq,~(\ref{eq:evolutionPhi}). This can be justified due to the fact that nuclei formation happens on a much shorter length (time) scales than the diffusion of solute. Hence, the inclusion of flux fluctuations in the composition field (or equivalently in chemical potential Eq.~(\ref{eq:evolutionmu})) should not have a significant effect on the nucleation process and can be ignored \cite{echebarria2010onset,karma1999phase}.  

Heterogeneous nucleation is implemented by assigning the coupling constant $\lambda$ (nominally proportional to $1/H_\alpha$) that emerges in the scaled phase field equations with a dual role, as follows: During the free growth kinetics stage of any grain in the system, $\lambda$ is used in the usual manner to assure convergence of the thin-interface defined by $\phi_\alpha$ onto the kinetics of the sharp interface model \cite{echebarria2004quantitative}. However, in the process of nucleation, which happens concurrently through a "ghost field" order parameter (explained further in the next sub-section), $\lambda$ represents a quenched in spatial field (labelled $\lambda_{\rm het}$), whose local value depends on local solute concentration in the undercooled liquid and inoculant contact angle according to
\begin{equation}
\label{eq:lambda_hetro}
\lambda_{\rm het}=\frac{1}{2}\,\frac{I\,W\,\Delta f_{\nu}}{\gamma\sqrt{h(\theta_c)}}
\end{equation}
where $W$ is the phase field "thin interface" interface used in simulations, and $I$ is a constant \cite{echebarria2004quantitative}. The expression for the bulk free energy change is given in terms of local composition of liquid $c$ according to
\begin{equation}
\label{eq:nucleation_DF}
\Delta f_{\nu} = \frac{RT_{\rm m}}{\Omega_{o}}\,(1-k)(c-c^{\rm eq}_{l}(T)),
\end{equation}
where $\Delta c=(c-c^{\rm eq}_{l}(T))$ is the difference between the undercooled liquid composition relative to its co-existence value at temperature $T$, and $R$, $T_{\rm m}$, $\Omega_{\rm o}$, $k$ are universal gas constant, melting temperature, molar volume, and the partition coefficient, respectively.  Eq.~(\ref{eq:lambda_hetro}) can be recast as
\begin{equation}
\label{eq:lambda_hetro2}
\lambda_{\rm het}(c,\theta_c)=\frac{15}{16}\frac{RT_{\rm m}}{\Omega_{o}}\,\frac{(1-k)\,\Delta c}{H_{\rm R}\,\sqrt{h(\theta_c)}}\bigg(\frac{W}{W_{\rm R}}\bigg),
\end{equation}
where $W_{\rm R} = \gamma/(IH_{\rm R})$ is the solid-liquid interface and $H_{\rm R}$ is the nucleation barrier. With the above information in hand,  a detailed numerical algorithm for implementing nucleation proceeds as described in the next sub-section.
%%
%%%
\subsection{Numerical implementation of nucleation algorithm}
Analogously to Ref.~\cite{K_shampur}, we introduce an auxiliary ``ghost" order parameter field that tracks fluctuations of an ideal liquid, and is maintained in a separate system, with dynamics governed by Eqs.~(\ref{eq:evolutionPhi}) and (\ref{eq:evolutionmu}), but interacts with the ``physical" order parameter grains in the main system of interest. Tracking heterogeneous nucleation in the ghost field, and transferring a post-critical ghost nucleus to the main system of interest proceeds by the following algorithm: 
\begin{enumerate}
    \item Possible nucleation sites (inoculants) are randomly positioned in the bulk liquid phase. The inoculants are approximated as circles (spheres in 3D).
    \item The values of nucleation barrier of inoculants are set randomly for a range of contact angles in interval $[5^{o},\,15^{o}]$, and inoculant size is set to $1 \mu {\rm m}$. These parameters correspond approximately to heterogeneous nucleation on $\rm{TiB}_{2}$ particles in very dilute AlSi12 alloy \cite{xi2019effect}.
    \item Fluctuations that lead to nucleated order parameters are affected in a separate system containing  a ``ghost field" order parameter, which interacts with the grains in the main simulation domain of interest, but not vice versa.
    \item If the thermal fluctuations of the ghost field within the footprint volume of an inoculant particle overcome the heterogeneous nucleation barrier, over a number of mesh pixels whose volume is greater than the critical nucleation size, the ghost field order parameter is assigned to (i.e added to) an order order parameter in the main system corresponding to one of the crystal orientation. The order parameter chosen to re-assign in the main system is chosen randomly from the possible orientations represented by the $\phi_\alpha$. Here, we work with three possible crystal orientations: $\phi_{1}=0, \phi_{2}=\pi/12, \phi_{3}=\pi/6$, each with respect to $<001>$ growth axis. It is noted that volume of the nucleated grain is subsequently be excluded from nucleating again in the ghost field domain due to the aforementioned interaction algorithm.
    \item Once a ghost field order parameter is nucleated and added to the main simulation domain, which contains post-critical solidifying crystals, the value of coupling constant $\lambda$ is maintained at a constant value maintained over the entire domain of interest, chosen so as to control the appropriate free-growth interface kinetics.
\end{enumerate}
It is noteworthy that, although the focus of this study is heterogeneous nucleation, a similar approach can be applied for homogeneous nucleation.
%%
%%%
\subsection{Bench-marking the phase field model with nucleation}
The validity of the phase-field model to predict
heterogeneous nucleation and to capture the phenomenon of CET was benchmarked through simulations of directional solidification of a dilute Al-Si alloy for a range of thermal gradient ($G$) and solidification rates ($V$). Phase-field simulations of solidification are performed on 2D domains using a C\texttt{++} finite difference adaptive mesh refinement (AMR) code that incorporates MPI parallelization \cite{greenwood2018quantitative}. This program performs the numerical integration of Eqs.~(\ref{eq:evolutionPhi}) and (\ref{eq:evolutionmu}) on a dynamic adaptive mesh.

\begin{figure}[htbp!]
\centering
\includegraphics[width=0.5\textwidth]{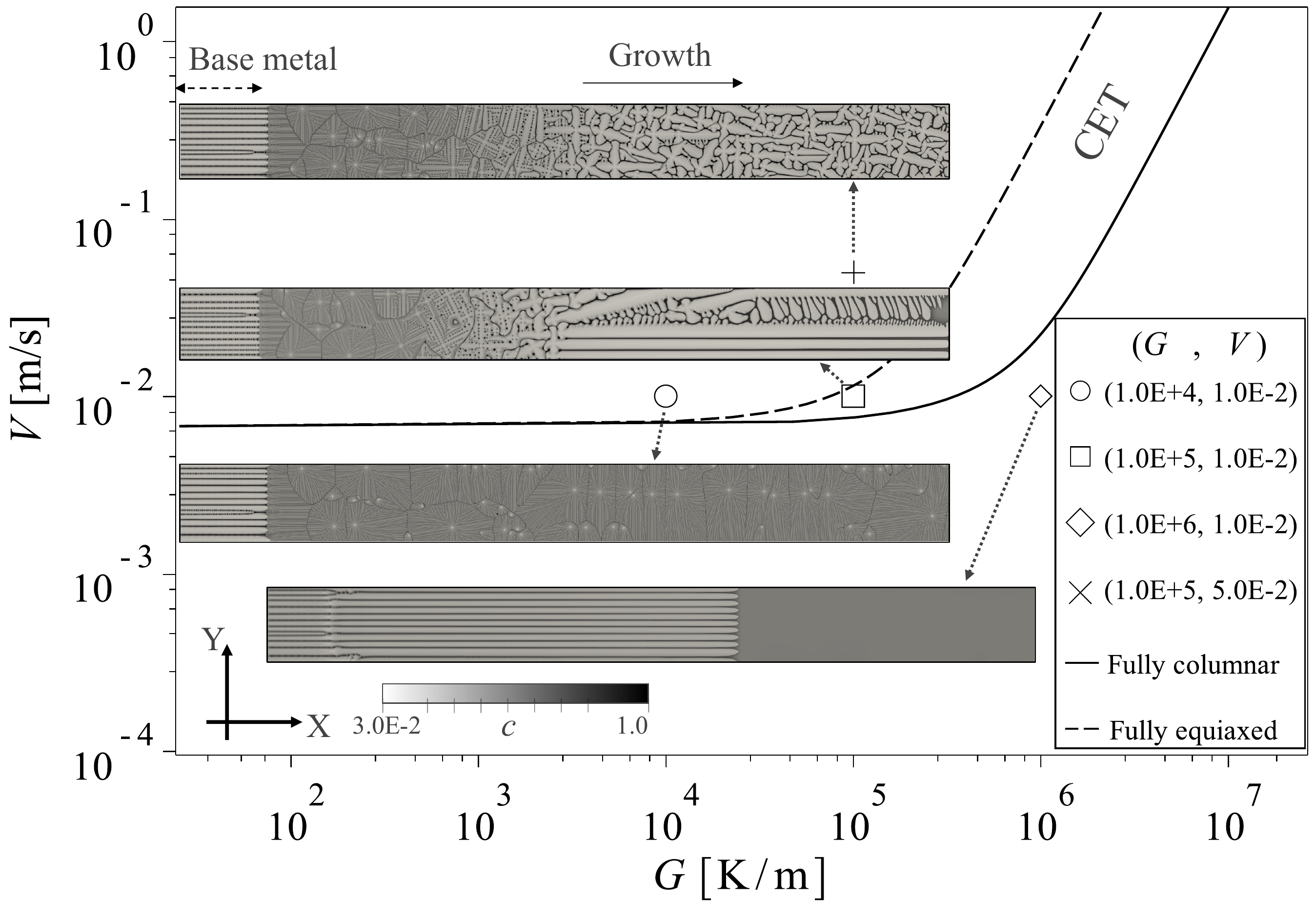}
\caption{(G-V) morphological phase diagram for dilute AlSi binary alloy using Hunt model. (Insets) solute composition map of the final solidification stage for phase field simulations for: ($G=10^{4} \mathrm{K/m},V=10^{-2} \mathrm{m/s}$,\protect\markersix),  ($G=10^{5} \mathrm{K/m},V=10^{-2} \mathrm{m/s}$, \protect\markerfour), ($G=10^{6} \mathrm{K/m},V=10^{-2} \mathrm{m/s}$, \protect\markerfive), and ($G=10^{5} \mathrm{K/m},V=5\times10^{-2} \mathrm{m/s}$, $\bigplus$). The analytical criteria for CET (i.e. fully columnar (\protect\solidrule) and fully equiaxed (\protect\dashedrule) curves) are calculated using Hunt steady-sate CET model for dilute AlSi alloy. The colour bar shows the dimension-less solute composition scale  (color online).}
\label{fig:2}  
\end{figure}

Simulations are performed within a 2D domain whose dimensions parallel and perpendicular to the growth direction are set to $X=300 \mu {\rm m} \times Y=30 \mu {\rm m}$, respectively. The initial conditions consist of a ``pre-solidified'' dilute AlSi alloy base metal. The  solidification process then initiates epitaxially from the base metal. This condition is meant to emulate a multi-layer deposit in the laser sintering processes. The nucleation site density and contact angle for heterogeneous nucleation are set to $N_{\rm p} = 2\times10^{15} \mathrm{m^{-3}}$ and $\theta_c=\pi/45$, respectively. The results of this numerical investigation for four pairs of $G,V$ parameters are presented in Fig.~\ref{fig:2}. The analytical criteria for columnar and equiaxed growth and their morphological transitions (i.e. CET) are calculated using a steady-state CET model developed by Hunt \cite{hunt1984steady} and are also presented in Fig.~\ref{fig:2}. The Hunt model is calibrated based on the nucleation criteria (i.e. nucleation rate and undercooling) that are extracted from the simulations. It is evident from these numerical results that increasing $G/V$ ratio corresponds to a transition from  equiaxed-dominated to columnar-dominated growth. This transition occurs either by increasing $G$ from ($G=10^{4} \mathrm{K/m}, V=10^{-2} \mathrm{m/s}$, \markersix) to ($G=10^{6} \mathrm{K/m}, V=10^{-2} \mathrm{m/s}$, \markerfive), or by decreasing solidification rate from ($G=10^{5} \mathrm{K/m}, V=5\times10^{-2} \mathrm{m/s}$, $\bigplus$) to ($G=10^{5} \mathrm{K/m}, V=10^{-2} \mathrm{m/s}$, \markerfour). The observed trend from our numerical results are consistent, qualitatively, with the analytical criteria. A small discrepancy observed between the numerical and analytical solutions originates from the simplifications made in the model by Hunt in describing the undercooling of dendritic growth. We expect a more quantitative agreement with the numerical model of Gaumann, Trivedi, Kurz (GTK)~\cite{gaumann1997nucleation} that incorporates local composition and local undercooling ahead of the moving solid-liquid interface.
%%
%%%%%%%%%%
\section{Thermal modelling}
The thermal profile of LPBF-AlSi10Mg for two horizontal and vertical printing strategies are calculated using the ABAQUS software package. 
In this study we employed the Lagrangian finite element analysis method \cite{abaqus20166}. 

The dimensions of the horizontal and vertical samples in ($X$,$Y$,$Z$) coordinates are set to $1\,{\rm cm} \times 1\,{\rm cm} \times 5\,{\rm mm}$ and $2.5\,{\rm mm} \times 2.5\,{\rm mm} \times 1\,{\rm cm}$, respectively.  In both cases the entire body is assumed to be already fabricated, with only the last powder layer of thickness $z=30\,\mu \rm m$ remaining to be solidified, which is the focus of our modelling. 

The spatio-temporal distribution of temperature in a 3D domain follows the energy transfer equation, and can be formulated as follows,
\begin{equation}
\label{eq:EnergyEq}
\rho\,c_{p}\,\frac{\partial{T(X,Y,Z,t)}}{\partial{t}}=-\nabla.\vec{q}(X,Y,Z,t)+Q(X,Y,Z,t),
\end{equation}
where $\rho$, $c_{p}$, and $T$ represent density, specific heat capacity at constant pressure, and temperature of the printed object, respectively. 
In Eq.~(\ref{eq:EnergyEq}) $\vec{q}=-k(T)\nabla{T}$ represents heat conduction for the material with temperature dependent heat conductivity $k(T)$, and $Q=Q_{\rm L}-Q_{\rm C}-Q_{\rm R}-{dH}/{dt}$ where $Q_{\rm L}$, $Q_{\rm C}$, $Q_{\rm R}$, and ${dH}/{dt}$ represent input heat flux of the laser, heat convection, heat radiation, and the release of latent heat, respectively.

The thermal heat source during LPBF is provided by a laser beam, whose flux distribution $Q_{\rm L}$ follows a Gaussian form, written as \cite{fu20143}, 
\begin{equation}
\label{eq:LaserGuassian}
Q_{\rm L}(r)=\frac{2AP}{\pi R^{2}}\,\exp\bigg(-\frac{2r^{2}}{R^2}\bigg),
\end{equation}
where $A$ is the laser absorptivity for the AlSi10Mg powder, $P$ is the laser power, $r$ is the radial distance from a point on the powder surface to the centre of the laser, and $R$ is the effective laser radius.

The heat flux for convection $Q_{\rm C}$, and radiation $Q_{\rm R}$ through open surfaces (i.e. exposed to air) are given by,
\begin{eqnarray}
Q_{\rm C}&=h\,(T-T_{\rm 0}) \nonumber \\ 
Q_{\rm R}&=\sigma\,\epsilon\,(T^{4}-T^{4}_{\rm 0}),
\label{eq:Convec_RadiaEqs}
\end{eqnarray}
where $h$ is the convection rate, $\sigma$ is the Stefan-Boltzmann constant, and $\epsilon$ is the emissivity.

The powder layer is represented as porous material, whose effective thermal conductivity $K_{\rm eff}$ and density $\rho_{\rm bed}$ are given by \cite{antony2014numerical},
\begin{eqnarray}
K_{\rm eff}&={\frac{\rho_{\rm R}\,K_{\rm s}}{1+\Phi\,\frac{K_{\rm s}}{K_{\rm g}}}} \nonumber \\ 
\rho_{\rm bed}&={\rho_{\rm R}\,\rho_{\rm s}=\frac{\pi}{6}\,\rho_{\rm s}}.
\label{eq:Effconduct_densitypowder}
\end{eqnarray}
In Eq.~(\ref{eq:Effconduct_densitypowder}), $K_{\rm s}$, and $\rho_{\rm s}$ are the thermal conductivity and density of material in the solidus state, respectively, $\rho_{\rm R}$ is the initial relative density, $K_{\rm g}$ is the thermal conductivity of the surrounding gaseous environment, and $\Phi$ is an empirical coefficient.  The LPBF process parameters used in this model are summarized in Table \ref{tab:1}.

In our simulation, the material begins as powder until the liquidus temperature, $597 ^\circ {\rm C}$ is reached. As the liquid phase becomes more undercooled during the cooling cycle, it undergoes a phase change to solid  as the temperature drops below the solidus temperature, $577 ^\circ {\rm C}$. If the initial material is found to have a temperature below the liquidus but has undergone the transformation to liquid, we assume it has solidus properties. The transformations from powder to liquid and liquid to solid phases are incorporated into the thermal model through the latent heat of fusion during phase change, which is modelled by expressing the enthalpy $H$ as a function of temperature $H=\int{\rho\,c_{p}\,dT}$.

\begin{figure*}[!t]
\centering
\includegraphics[scale=.7]{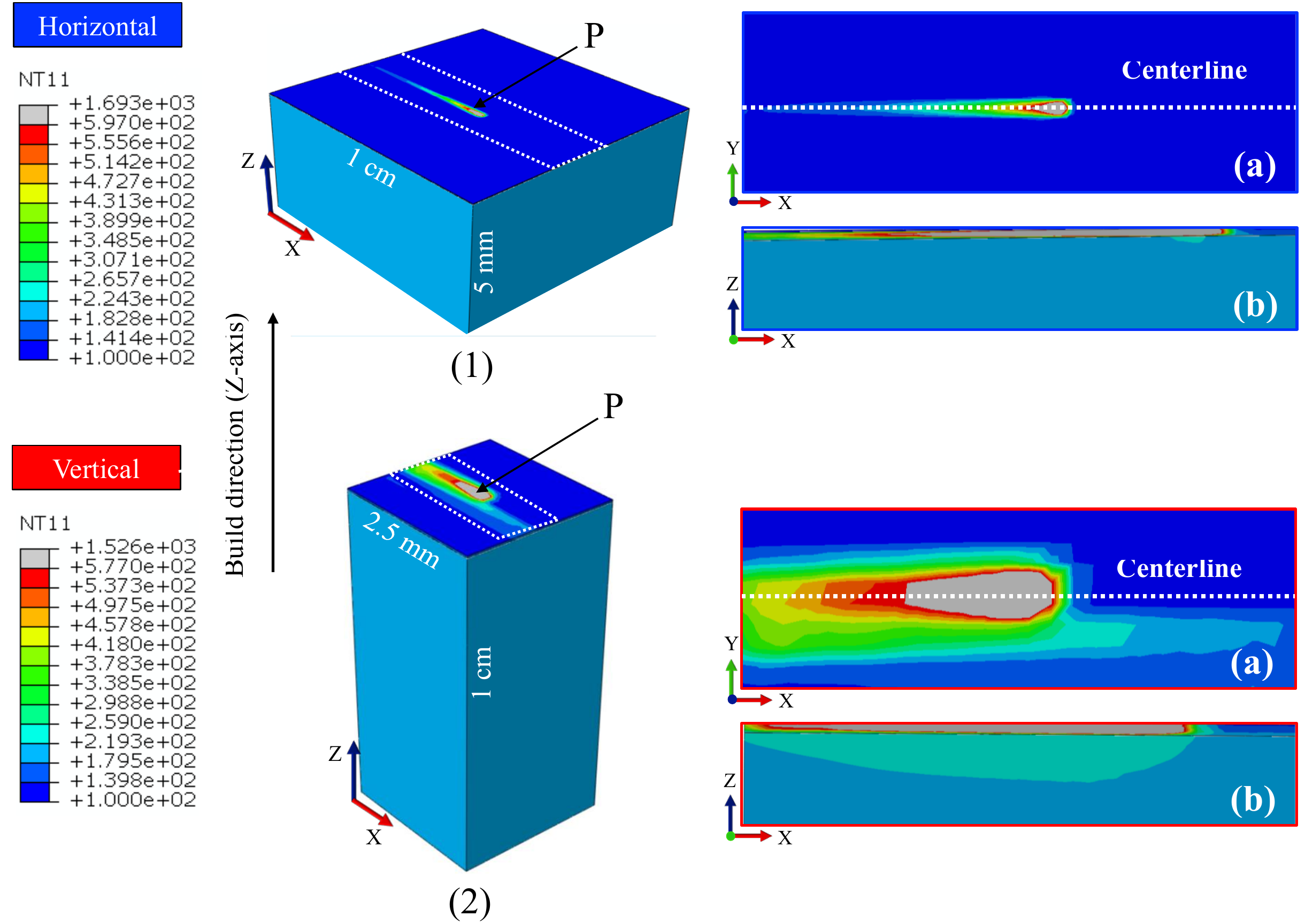}
\caption{The geometry of the printed layer with (1) horizontal and (2) vertical build orientations and resulting thermal profiles (a) on the top surface (X-Y plane), and (b) on the cross-section (X-Z plane) of the molten pool. The position of the laser is shown by \textbf{P}. The colour bar shows the temperature scale in Celsius. The molten pool is depicted by gray colour.}
\label{fig:3}  
\end{figure*}

The geometry of the horizontally and vertically printed layer with their resulting thermal profile on the top surface (X-Y plane) and cross-section (X-Z plane) of the molten pool are depicted in Fig.~\ref{fig:3}. These results show that both length ($L_{\rm X}$) and width ($L_{\rm Y}$) of the molten pool (shown by the gray colour) on the top surface, in the vertical sample are considerably larger than those in the horizontal one.  This can be attributed to a shorter time period between each two successive scanning tracks passing a certain point in the vertical sample compared to the horizontal case, which effectively increases the melt pool dimensions and liquid life time. On the other hand, the more effective heat dissipation mechanisms in the horizontal sample can lower temperatures between two consecutive scanning tracks, giving rise to higher cooling rates. 

As evident in Fig.~\ref{fig:3}(1-2)(b), the depth of the molten pool ($L_{\rm Z}$) is almost identical for both samples. This suggests that the build orientation does not seem to have considerable impact on the molten pool depth. As shown in previous studies the penetration depth of the laser is mainly controlled by the laser energy density ($\propto$ Power/Scan speed)~\cite{yu2016influence}. The maximum depth of the melt pool exceeds the powder layer and reaches $\approx 36 \mu {\rm m}$, which indicates that the laser beam can penetrate into the preceding layer, creating a maximum re-melting depth of $6 \mu {\rm m}$. The relatively small penetration depth can be attributed to the large scanning speed, which at a fixed laser power effectively lowers the laser energy per unit length \cite{yu2016influence}.

\begin{table}
\caption{Processing parameters of LPBF-AlSi10Mg.}
\label{tab:1}       % Give a unique label
\begin{tabular}{p{4.7cm}p{4cm}}
\hline\noalign{\smallskip}
Scan speed, $V_{B}$ ($\rm ms^{-1}$) 	& \hspace{75pt} 1.3 	\\
Laser Power, $P$ ($\rm W$)  	& 	\hspace{75pt} 370\\
Laser spot size, $R$ ($\rm mm$) & \hspace{75pt} 0.1  \\
Layer thickness ($\rm mm$) & \hspace{75pt} 0.03 \\
Preheat temperature, $T_{\rm o}$ ($\rm K$)& \hspace{75pt} 473.15 \\
Thermal diffusivity ($\rm m^{2}s^{-1}$)& \hspace{75pt} $4\times10^{-5}$ \\
Heat transfer, $h$ ($\rm W\,m^{-2}\,K^{-1}$)& \hspace{75pt} 80 \\
Emissivity, $\epsilon$ & \hspace{75pt}  $0.3$  \\
Powder absorptivity, $A$ & \hspace{75pt}  $18 \%$  \\
K$_{\rm g}$ & \hspace{75pt}  $0.024$ \\
$\Phi$ & \hspace{40pt} $0.02\times10^{2(0.7-\rho_{\rm R})}$ \\
\noalign{\smallskip}\noalign{\smallskip}
\hline\noalign{\smallskip}
\end{tabular}\\
\vspace*{-10pt}
\end{table}
%%

%%%%%%%%%
\section{Results and Discussions}
To examine the effect of building orientation on the microstructure of LPBF-AlSi binary alloy, we conduct phase-field simulations of the cross-section of the molten pool (i.e X-Z plane through the centerline) of a vertically and horizontally sintered single powder layer (Fig.~\ref{fig:3}(1-2)(b)). The main difference between horizontal and vertical build directions, as discussed in the preceding section, is that the laser passes in the vertical sample are considerably shorter than the horizontal one, which results in higher residual heat and larger melt pool dimensions in the former. The simulation domain for both horizontal and vertical samples consists of a 2D rectangular domain ($L_{\rm X}=400\,\mu \rm m$ , $L_{\rm Z}= 40\, \mu \rm m$), where the spatial coordinates $X$ and $Z$ are parallel and normal to the direction of laser propagation, respectively. All relevant parameters for the simulation is summarized in table \ref{tab:2}. 

\begin{figure*}[t!]
\includegraphics[scale=0.4]{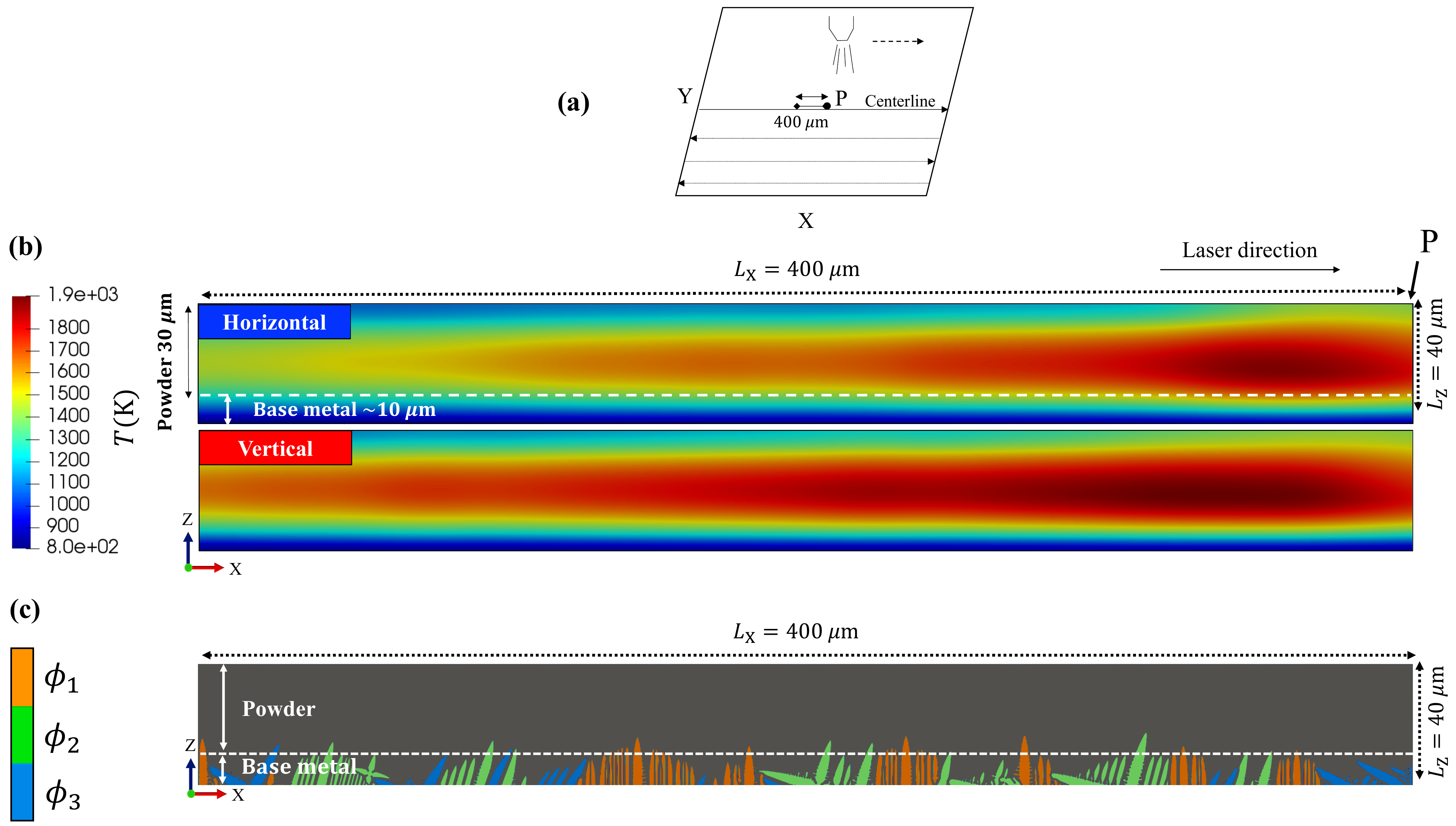}
\centering
\caption{a) Laser scanning pattern during LPBF process (point P: position of the laser at the center of the layer along the molten pool centerline). b) Temperature distribution of the cross-section of the molten pool where the laser is above point P. The colour bar shows the temperature scale in Kelvin. c) The single-layer deposit and base metal system. The base metal is composed of several solid grains with random crystal orientations, with respect to $[001]$ direction, which are depicted by \textcolor{orange}{$\phi_1=0$}, \textcolor{limegreen}{ $\phi_2=\pi/12$}, and \textcolor{blue}{$\phi_3=\pi/6$.} }
\label{fig:4}  
\end{figure*}

The single-layer deposit system used in this study comprises a powder layer ($30 \mu{\rm m}$) distributed evenly on top of a base metal ($10 \mu{\rm m}$) as depicted in Fig.~{\ref{fig:4}(b)}. The base metal comprises of several solid grains with random crystal orientations (shown by different colors in the figure) and represents a fraction of the preceding layer. 

We neglect the porous nature of the powder layer and other features such as densification that might occur as a result. We expect that this is adequate for our purposes and prove this assumption moving forward.  

The powder particles are mostly melted due to the high temperature of the laser, and thus they do not seem to have a significant effect on the resultant microstructure. 

\begin{figure*}[t!]
\centering
\includegraphics[scale=0.45]{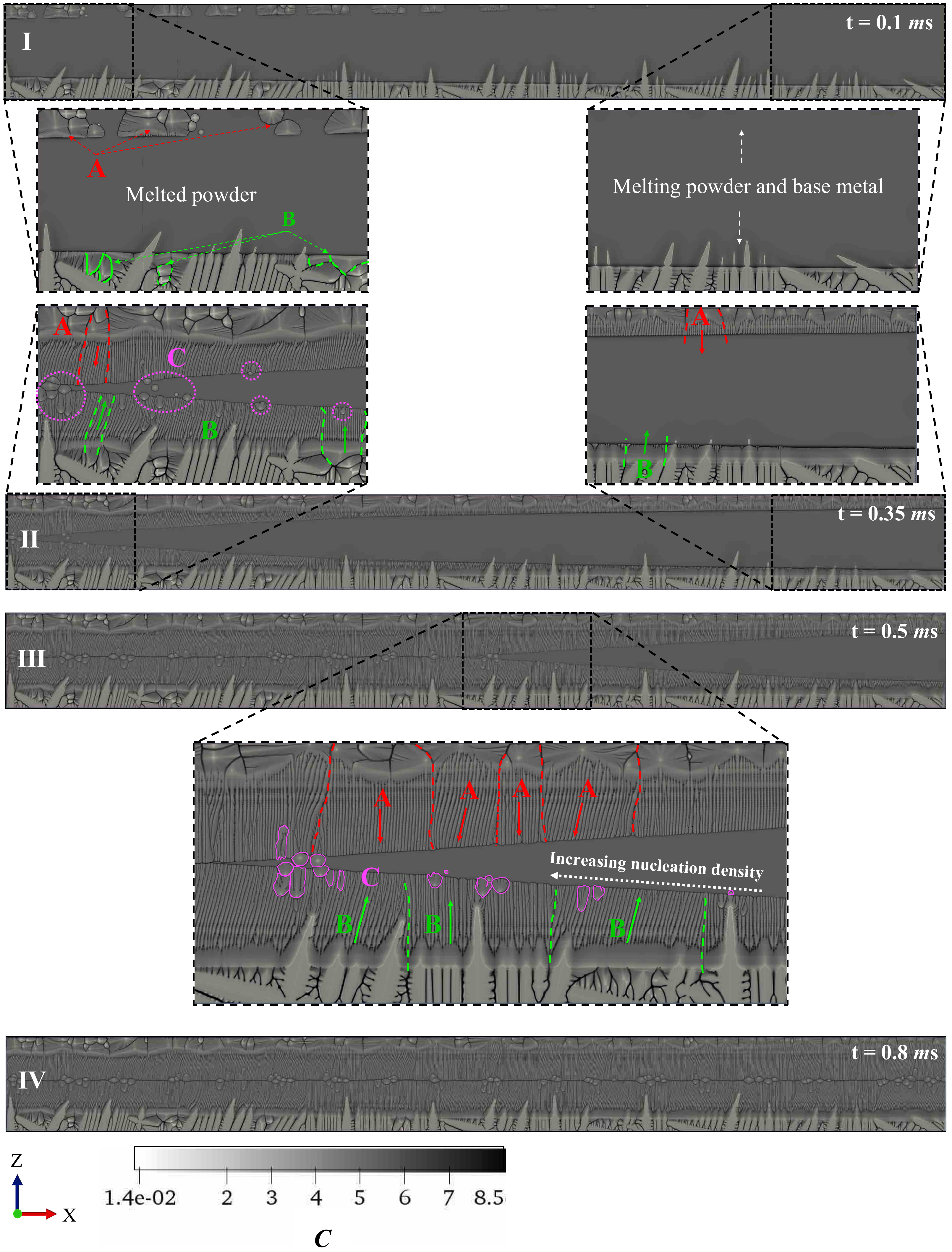}
\caption{The sequence of time evolution of concentration map of solute (Si) (microstructure) in the sintered single powder layer with horizontal build direction: \RNum{1}) after initial nucleation events, \RNum{2}) at early stage of solidification, \RNum{3}) an intermediate solidification stage (almost half of the molten pool is solidified), and \RNum{4}) after complete solidification. Three distinct types of grain growth are designated by A-C (see text for details). Time is in units of $\times10^{-3} \rm {s}$. The colour bar shows the dimensionless solute concentration scale. Coloured dashed lines (arrows) in the insets are used to specify distinct crystal orientation (growth direction) of {\color{red}{A}} and {\color{limegreen}{B}} columnar grains.}
\label{fig:5}  
\end{figure*}

We initiate our simulations at $t=0$ by imposing the thermal profile of the molten pool on the powder-base metal system, when the laser position is above the point {\textbf{P}} (Fig.~{\ref{fig:4}(a),(b)}). The powder starts to gradually melt under the imposed thermal conditions, which creates a molten pool. The melt pool boundary (i.e. solid-liquid interface) then evolves according to the corresponding transient laser thermal profile. The results of these simulations are depicted in Fig.~\ref{fig:5}. For the sake of brevity, we only elaborate the sequence of time evolution of microstructure in the horizontally sintered powder. The same discussion can be applied to the vertical sample.   

Figure~\ref{fig:5} shows the solute (Si) concentration map for four instances during solidification for the horizontal sample. The instances correspond to; \RNum{1}: after few initial nucleation events on the top surface; \RNum{2}: at early solidification stage; \RNum{3}: an intermediate solidification stage (almost half of the domain is solidified); and \RNum{4}: after complete solidification. We classify different ``types'' of grains observed in the horizontal and vertical samples by designating them as type A-C in the insets of Fig.~\ref{fig:5}.

At the early stages of the simulation several nuclei have emerged on the top surface with random orientations (marked by A), which continue to grow, driven by the local thermal gradient. The formation of these nuclei can be attributed to the relatively large heat dissipation rate in the vicinity of top surface, which give rise to high cooling rates ($\dot{T}=\partial{T}/\partial{t}$).

These grains collectively form a secondary, downward propagating (solid-liquid) interface (DPI)(Fig.~{\ref{fig:5}}(\RNum{2})). Among all the grains with different orientations that nucleated initially, only those aligned similarly with the local temperature gradient tend to grow fastest and thus outgrow slower misaligned grains.

Figure~{\ref{fig:5}}(\RNum{1}),(\RNum{2}) also shows that an upward propagating interface (UPI) is formed at the bottom, which consists of fine columnar grains (B grains) growing either on the top of base metal grains. This can be considered as an epitaxial growth, or from initially nucleated equiaxed grains in between the base metal dendrites. Emergence of these fine grains from much coarser base metal dendrites is indicative of grain growth under high cooling rate conditions. This was observed in an experimental study of  LPBF-AlSi10Mg on an Aluminum-based cast alloy substrate \cite{hadadzadeh2020microstructural}. Unlike the top and bottom ends of the domain, the middle part undergoes a prolonged re-melting stage ( Fig.~{\ref{fig:5}}(\RNum{1}),(\RNum{2})). 

At later time, both the DPI and UPI merge at the trailing end of the melt pool shaping a zipper-like structure that forms a cohesive solid (Fig.~{\ref{fig:5}}(\RNum{2})).  The zipper structure is the direct consequence of the thermal boundary and cooling conditions during the sintering process of the single powder layer, and is expected to appear in the skin layer of fabricated parts. It is worth noting that the zipper structure reported here is similar to those that are observed in an experimental study on microstructural characteristics of stainless-steel 304L parts produced using directed energy deposition (DED) process \cite{wang2016effect}.

The imposed cooling conditions on the top surface that establish large thermal gradient $G$ across the surface (through convection and radiation), can also explain larger solidification rate and thus larger grain sizes on the top (A grains) compared to the bottom (B grains) of the molten pool (Fig.~{\ref{fig:5}}(\RNum{2})).

As evident from Fig.~{\ref{fig:5}}, at this stage, a few heterogeneous nucleation events (C grains in Fig.~{\ref{fig:5}}) also emerge in the undercooled region between the two approaching fronts. Some nucleated events are clearly displayed in the inset of Fig.~{\ref{fig:5}}(\RNum{2})).

The trailing (solidifying) part of the melt pool in the intermediate solidification stage is illustrated in the inset of Fig.~{\ref{fig:5}}(\RNum{3}). As shown in the figure, the columnar B grains have the same crystallographic orientation as the seed crystal (i.e. base metal dendrites in this case or the nucleated equiaxed grains as displayed in the lower-right part of the top-left inset of Fig.~{\ref{fig:5}}(\RNum{2})) \cite{rappaz1989development}. 

The situation for A grains is different from B grains as there are initially multiple nucleated equiaxed grains within a small surface element around a given location with random orientations.  The growing direction of the final dendrite from the location will be along the direction of the nucleated crystal that is most closely aligned with the local gradient vector \cite{liu2017quantitative}.

The nucleation density of C grains increases rapidly along the melt pool boundary from leading to the trailing edge, which signals the higher undercooling in these areas. The solidification regime at a given location along the melt pool depends on the local $G$ and $V$, which not only depends on the process parameters but also vary along the melt pool. The grain boundaries at the trailing part of the melt pool, where two interfaces merge, tend to be sloped in the direction of beam travel (i.e. X-axis), which can alter both solidification rate and thermal gradient, thereby enhance the undercooling and thus nucleation rate.

\begin{figure*}[t!]
\centering
\includegraphics[scale=0.4]{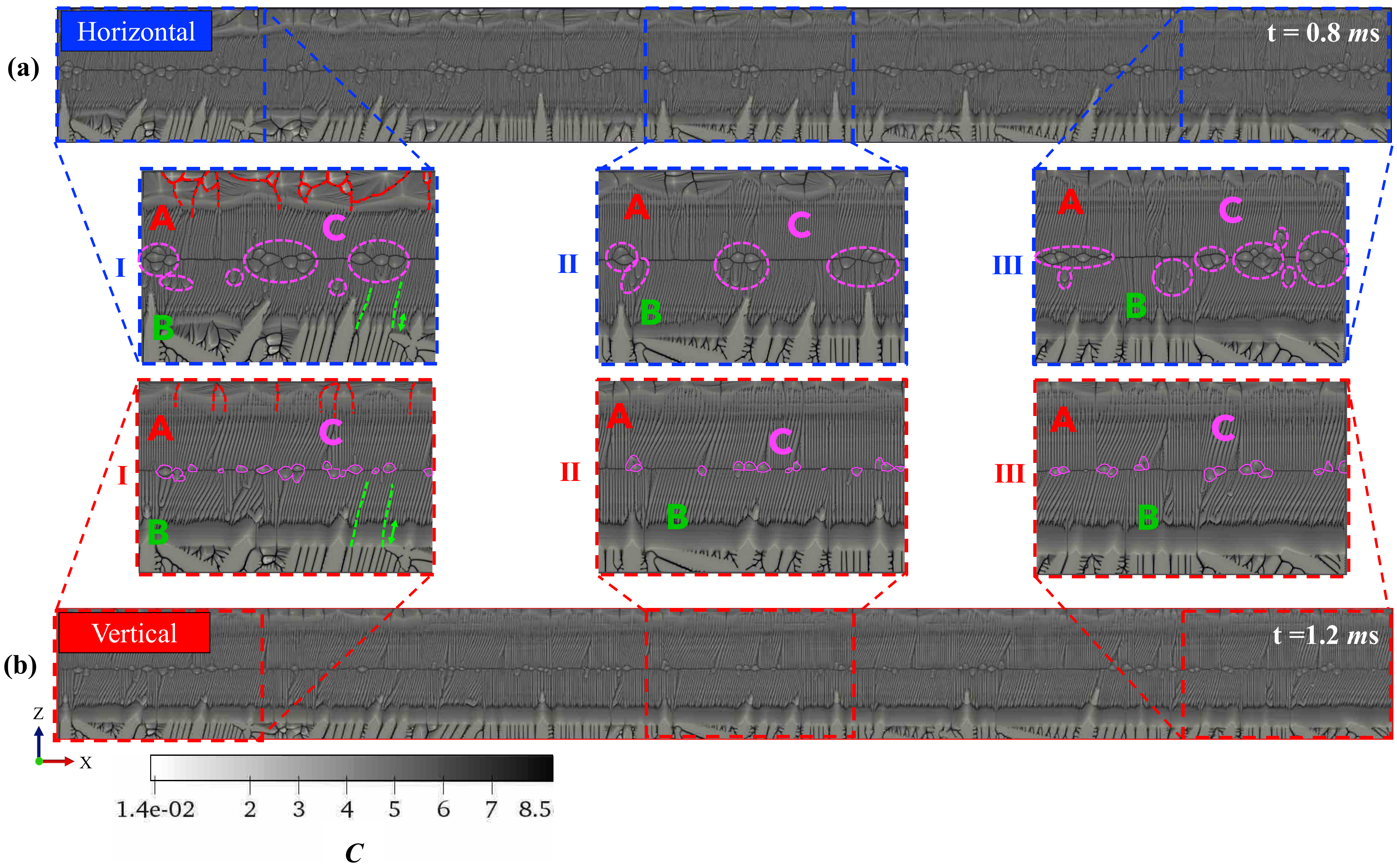}
\caption{The sequence of time evolution of concentration map of solute (Si) (microstructure) in the fully solidified sintered single powder layer with horizontal (top) and vertical (bottom) build direction: insets show the grain structure at the \RNum{1}) beginning, \RNum{2}) middle, and \RNum{3}) end of the domain.  Three distinct types of grain growth are marked by A-C. Time is in units of $\times10^{-3} \rm {s}$. The colour bar shows the dimensionless solute concentration scale.}
\label{fig:6}  
\end{figure*}

The final stage of the simulations are displayed in (Fig.~{\ref{fig:5}}(\RNum{4})). The grain structures of the fully solidified horizontally and vertically fabricated powder layer are better illustrated in the insets of Fig.~{\ref{fig:6}}.

Our results clearly suggest that the number of equiaxed grains (formed by heterogeneous nucleation type A and C) is significantly larger in the horizontal sample (see Fig.~{\ref{fig:6}}(a-b)(\RNum{1})). This can be attributed to the constitutional underccoling stability which is characterized by $\frac{G}{V}\geq\frac{m_l\,c^{eq}_l(1-k_{e})}{D_l}$, where $c^{eq}_l$ is the equilibrium liquids solute (Si) composition at solid-liquid interface \cite{tiller1953redistribution}.

Owing to its local cooling conditions and thermal boundaries, curvature of the trailing part of melt pool boundary is more pronounced in the horizontal sample compared to the vertical sample, which effectively lowers $\frac{G}{V}$ and creates a larger constitutional undercooling in the vicinity of the solid-liquid interface. Thus, the larger undercooling  leads to a larger number of nucleation events in the former sample. Nonetheless, the nucleation density in either case, is much less than the density of inoculants. A possible explanation for this behaviour is proposed next. 

A relatively large thermal gradient, $G$, along with fast solidification rates, $V$ in the LPBF process, can give rise to very high cooling rates $CR=G\times V$ ($10^{5}-10^{6}\,\mathrm{K/s}$) for the most part of the molten pool which effectively reduces constitutional undercooling and consequently suppresses the probability of nucleation events.  Furthermore, as the solidification front progresses, it rejects solute into the liquid which increases the solute concentration in the vicinity of solid-liquid interface. This results in the liquid in between the two approaching interfaces becoming highly saturated ( i.e. ($c-c^{\rm eq}_{l}(T))\to 0$ in Eq.~(\ref{eq:nucleation_DF})) and therefore further nucleation of equiaxed grains in these regions becomes very improbable.

As evident from Fig.~{\ref{fig:6}}(a-b)(\RNum{1}), columnar B grains for both samples undergo a short-lived dendritic to planar transition. This transition corresponds to the transition between remelting and solidification of the base metal during which $G$ is maximum and $V$ is minimum (which maximize $\frac{G}{V}$). The solidification parameters during this transition falls in the stable planar front region of the $G-V$ morphological phase diagram obtained by Dantzig et al. \cite{dantzig2016solidification}. As the solidification proceeds cooling conditions at the melt pool boundary tend to decrease $G$ and increase $V$. The planar interface, thus become unstable and evolves into columnar grains. The planar to columnar transition rate is more prolonged in the vertical case due to the smaller cooling rates relative to the horizontal sample.

Our simulation results of the effect of building direction on the microstructure evolution are consistent qualitatively with the experimental observations (Fig.~\ref{fig:1}). A quantitative discrepancy observed is due the fact that the EBSD in Fig.~\ref{fig:1} come from to an arbitrary location from the interior of the horizontal and vertical samples, whereas our simulation specifically shows the grain structure on the last deposit layer of the fabricated part.  Nevertheless, We expect to observe a similar zipper structure if a single deposit layer were to be characterized. 

To further examine the effect of build orientation on solidification regime, and eliminate the solute contamination effect from the DPI, we isolate the microstructure evolution for the UPI by only considering lower half of the thermal profile. The results of these analysis will be presented in the next section.
\begin{table}
\caption{Physical properties of AlSi binary alloy, phase-field and nucleation parameters used in the simulations.}
\label{tab:2}       % Give a unique label
\begin{tabular}{p{4.7cm}p{4cm}}
\hline\noalign{\smallskip}
{\bf{Physical properties}} &   \\
\noalign{\smallskip}
Chemical composition (wt. \%)&  	\hspace{45pt} Al:99.5,Si:0.5 	\\
Melting point of Al ($\rm K$)  	& 	\hspace{65pt} 932.85\\
Liquids slope ($\rm ^{o}C\,wt.\%^{-1}$) & \hspace{65pt} 6.5  \\
Partition coefficient, $k_{\rm e}$& \hspace{65pt} 0.13 \\
Solute diffusivity (liquid; $\rm m^{2}s^{-1}$)& \hspace{65pt} $3\times10^{-9}$ \\
Solute diffusivity (solid; $\rm m^{2}s^{-1}$)& \hspace{65pt} $1\times10^{-12}$ \\
Gibbs-Thomson coefficient($\rm K\,m$)& \hspace{65pt} $9\times10^{-8}$  \\
Density ($\rm kg\,m^{-3}$) &   \hspace{65pt} 2650\\
Latent heat ($\rm J\,kg^{-1}$) & \hspace{65pt} 389187 \\
%Density ($\rm kg\,m^{-3}$)$^a$ &   \hspace{75pt} 2550\\
\noalign{\smallskip}
{\bf{Phase-field parameters}} & \\
\noalign{\smallskip}
Effective interface width, $W_{\rm{o}}$ ($\rm m$)& \hspace{65pt} $2.5\times10^{-8}$\\
Relaxation time, $\tau_{\rm o}$ ($\rm s$)& \hspace{65pt} $5.7\times10^{-7}$\\
Minimum grid spacing, $\rm dx$& \hspace{65pt} $0.8\,W_{\rm o}$\\
Anti-trapping coefficient & \hspace{65pt} 0.35355\\
\noalign{\smallskip}
{\bf{Nucleation parameters}} & \\
\noalign{\smallskip}
Inoculant density, $N_{\rm{p}}$ ($\rm m^{-3}$)& \hspace{65pt} $2\times10^{15}$\\
Contact angle, $\theta_c$ & \hspace{64pt} $[5^{o},\,15^{o}]$\\
\noalign{\smallskip}
\hline\noalign{\smallskip}
\end{tabular}\\
\vspace*{-10pt}
\end{table}
%%

%%%%%%%%%%%
\subsection{Microstructure evolution of half-powder base metal system}
\begin{figure*}[t!]
\centering
\includegraphics[scale=0.4]{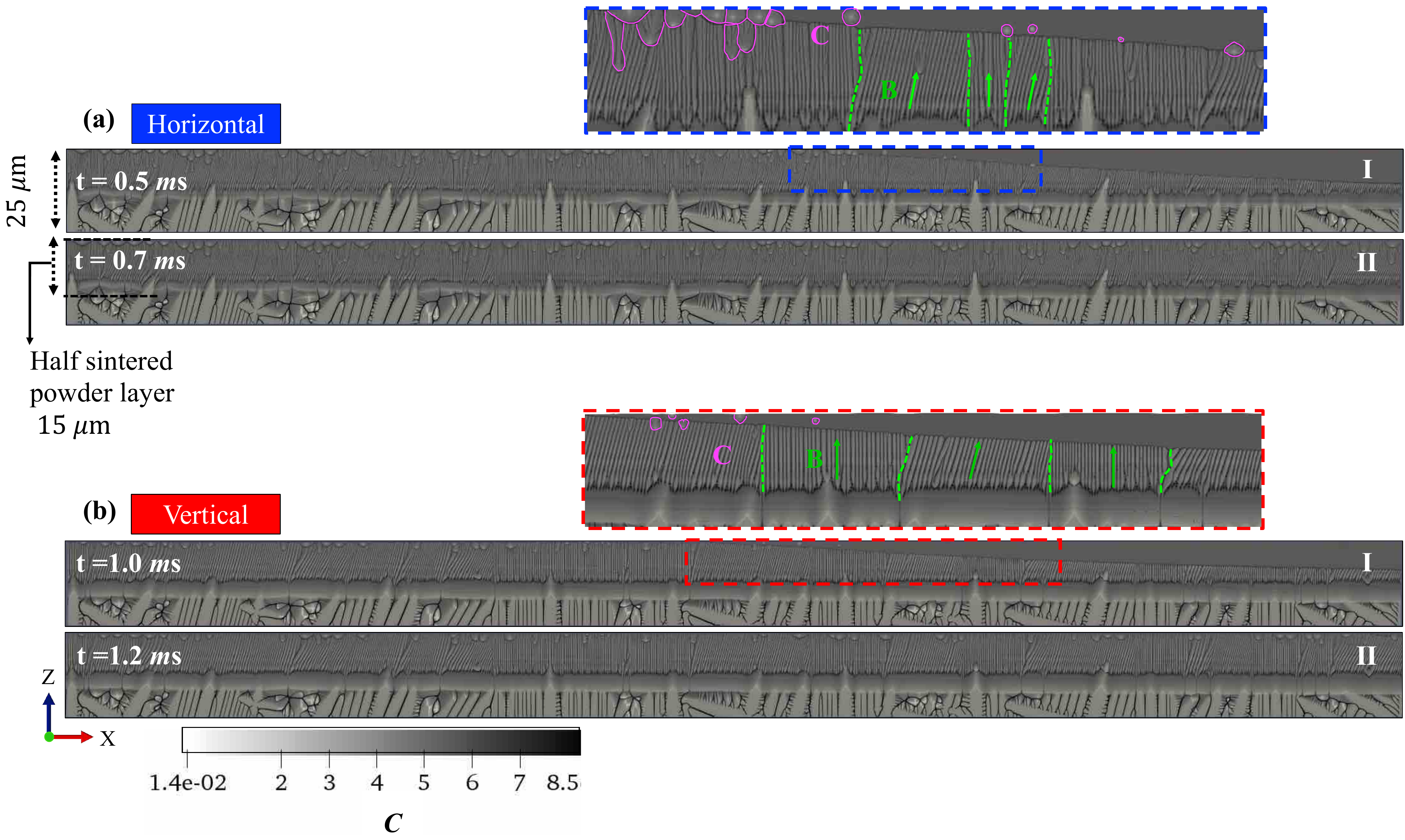}
\caption{The sequence of time evolution of concentration map of solute (Si) (microstructure) in the half sintered single powder layer with horizontal (top) and vertical (bottom) build direction: \RNum{1}) an intermediate solidification stage (almost half of the molten pool is solidified), and \RNum{2}) after complete solidification. Two distinct types of grain growth are marked by B and C (see text for details). The colour bar shows the dimensionless solute concentration scale.}
\label{fig:7}  
\end{figure*}

We conducted the same phase-field simulations of grain evolution in the cross-section of  the  molten pool as explained in the preceding section, except only the lower half of the molten pool (i.e. $15 \mu \mathrm{m}$) is considered, geometrically and thermally. The evolution of grain structure for both build orientations are displayed in Fig.~\ref{fig:7}.  

The microstructure in both build directions are composed of equiaxed grains C nucleated in the undercooled region ahead of propagating columnar B dendrites.  The number of nucleated equiaxed grains, analogously to the the data shown previously in Fig.~\ref{fig:6}, is higher in the horizontal sample compared to the vertical one. 

The difference between grain structures of the fully solidified layer for the horizontal and vertical build directions  (Fig.~{\ref{fig:7}}(a-\RNum{2}), (b-\RNum{2})) is better seen in in the data of Fig.~{\ref{fig:8}}. In Fig.~{\ref{fig:8}} three distinct crystal orientations are shown with respect to the major axis for FCC alloys $[001]$ (along Z-axis), represented by (\textcolor{orange}{$\phi_{1}=0$}), (\textcolor{limegreen}{$\phi_{2}=\pi/12$}), and ({\textcolor{blue} {$\phi_{3}=\pi/6$}). Different shades of the colors depicts the distribution of solute (Si) composition.    
\begin{figure*}[t!]
\centering
\includegraphics[scale=0.4]{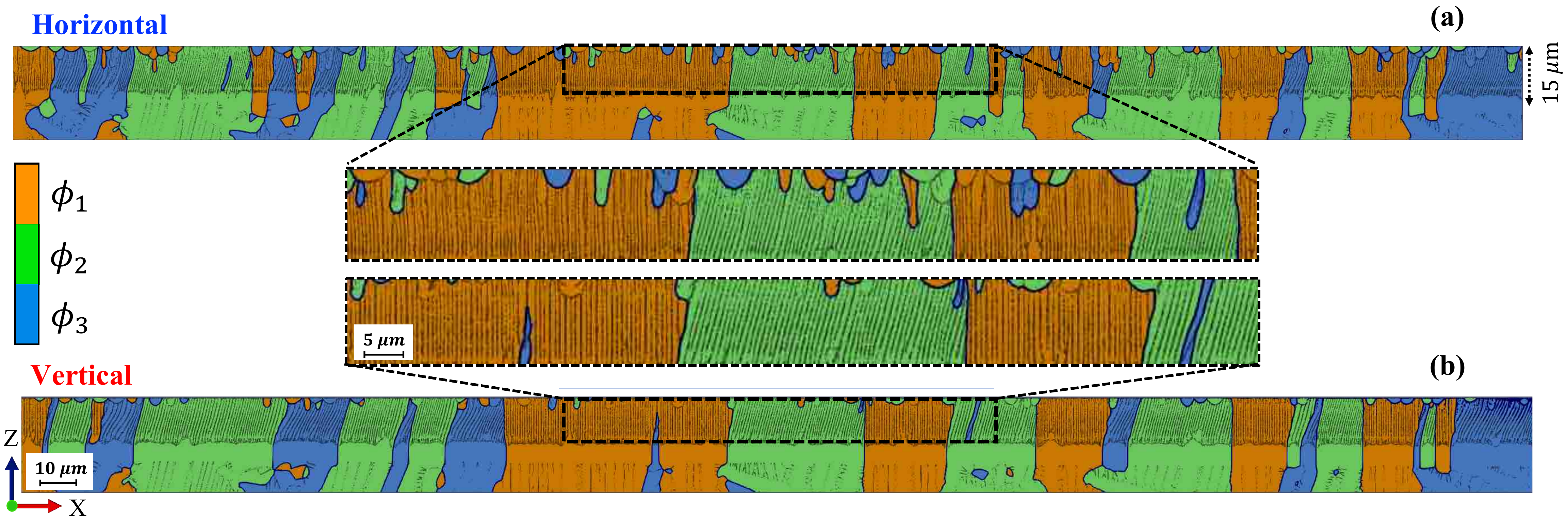}
\caption{Grain structures of the sintered half powder layer with a) horizontal and (b) vertical build directions. Colours represent three crystal orientations \textcolor{orange}{orange} ($\phi_{1}=0$), \textcolor{limegreen}{green} ($\phi_{2}=\pi/12$), and \textcolor{blue}{blue} ($\phi_{3}=\pi/6$). }
\label{fig:8}  
\end{figure*}

The distinction between the resultant microstructure for different build orientation in the data of Figs.~{\ref{fig:7}} and {\ref{fig:8}}  can be explained based on the constitutional undercooling for each case \cite{tiller1953redistribution}. 
The constitutional undercooling is defined as $\Delta T=T_{E}-T_{l}$, where $T_{E}$ is the equilibrium liquids temperature, given by  
\begin{equation}
T_{E}(z)=T_{\rm m}-m_{l}\,c_{l}(z),
\label{eq:Equitemp}
\end{equation}
where $c_{l}(z)$ is the liquids solute (Si) composition corresponding to a position $z$ (relative to the interface),  written for a steady state moving interface as
\begin{equation}
c_{l}(z)=\bigg[c_{\rm o}+\big(c^{\rm eq}_{l}(T_{\rm int})-c_{\rm o}\big)\exp\bigg(\frac{-V}{D_{l}}z\bigg)\bigg],
\label{eq:Equi_composition}
\end{equation}
and where $T_{l}$ is the local temperature of the molten pool within a small region in the vicinity of the solid-liquid interface, which is approximated by
\begin{equation}
T_{l}(z)=T_{\rm int}+Gz.
\label{eq:localtemp}
\end{equation}
In Eqs.~(\ref{eq:Equitemp})-(\ref{eq:localtemp}),
$T_{\rm int}$ and $c^{\rm eq}_{l}(T_{\rm int})$ are temperature and equilibrium liquids composition at solid-liquid interface, $c_{\rm o}$ is the average alloy composition, and $z$ is the distance from the solid-liquid interface. 
By employing Eqs.~(\ref{eq:Equitemp})-(\ref{eq:localtemp}), the undercooling $\Delta T$ for local velocity $V$ and thermal gradient $G$ at the interface can be recast as,
\begin{multline}
\label{eq:constundercooling}
 \Delta{T(z)} = T_{m}-\bigg(T_{\rm int}+Gz\bigg)- \\
+ m_{l}\bigg[c_{\rm o}+\big(c^{\rm eq}_{l}(T_{\rm int})-c_{\rm o}\big)\exp\bigg(\frac{-V}{D_{l}}z\bigg)\bigg)\bigg].
\end{multline}
To better represent the dependence of undercooling in Eq.~(\ref{eq:constundercooling}) on the processing and materials parameters, we consider the maximum undercooling $\Delta{T_{\rm max}}$, which is found by solving $(\partial{\Delta{T(z)}}/\partial{z})\big|{_{z_{\rm max}}}=0$ for $z_{\rm max}$, which gives
\begin{equation}
\label{eq:zmax}
{z_{\rm max}}=-\frac{D_l}{V}\ln{\bigg[\frac{GD_l}{m_l\big(c^{\rm eq}_{l}(T_{\rm int})-c_{\rm o}\big)\,V}}\bigg].
\end{equation}
Substituting Eq.~(\ref{eq:zmax}) into Eq.~(\ref{eq:constundercooling}) gives $\Delta{T_{\rm max}}$ as,
% %%
\begin{multline}
\label{eq:DTmax}
 \Delta T_{\rm max}=T_{m}-T_{\rm int}-m_{l}\bigg(c_{\rm o}+\frac{G}{V}\frac{D_{l}}{m_{l}}\bigg) + \\
 \frac{G}{V}D_{l}\ln{\bigg[\frac{G}{V}\frac{D_{l}}{m_{l}\,(c^{\rm eq}_{l}(T_{\rm int})-c_{
\rm o})}\bigg].}
\end{multline}
We next require to estimate the local thermal gradient $G$ and interface (front) velocity $V$. 

\begin{figure*}[t!]
\centering
\includegraphics[scale=.37]{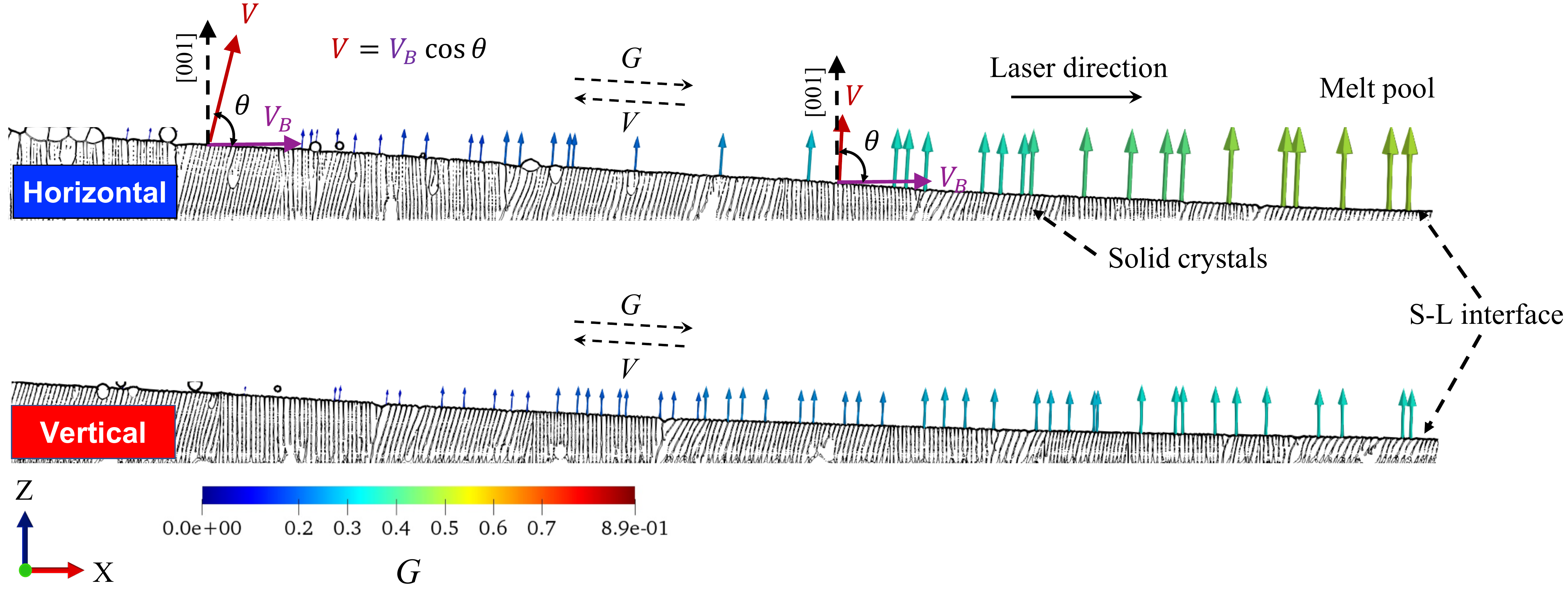}
\caption{Schematic representation of the local dendritic growth rate $V$ (dark red vectors), and thermal gradient $\vec{G}$ (coloured vectors) along the molten pool boundary for the horizontal (top) and vertical (bottom) build directions . The colour bar shows the dimensionless scale of the magnitude of thermal gradient vector. The dashed arrows show the direction of increase of $G$ and $V$.}
\label{fig:9}  
\end{figure*}

The thermal gradient $G=\left |{\nabla{T}}\right |$ at given time and position is calculated from the transient thermal profile of the molten pool (Fig.~\ref{fig:4}(b)). Moreover, the local solidification rate is related to the beam velocity $V_B$ and melt pool morphology, and can be approximated as $V=V_B\cos{\theta}$, where $\theta$ is the angle between direction of the moving source and growth direction of the solidifying material. The schematic representation of the relationship between laser speed ($V_{B}$), and the dendritic front growth rate $V$, along with the thermal gradient $G$ for both build directions are displayed in Fig.~{\ref{fig:9}.

As shown in the figure, the variation of $G$ and $V$ along the melt pool for both build directions follow the same trend where $G$ increases while $V$ decreases from the trailing to the leading edge of the melt pool (Fig.~{\ref{fig:9}}).
\begin{figure}[htb!]
\centering
\includegraphics[scale=.32]{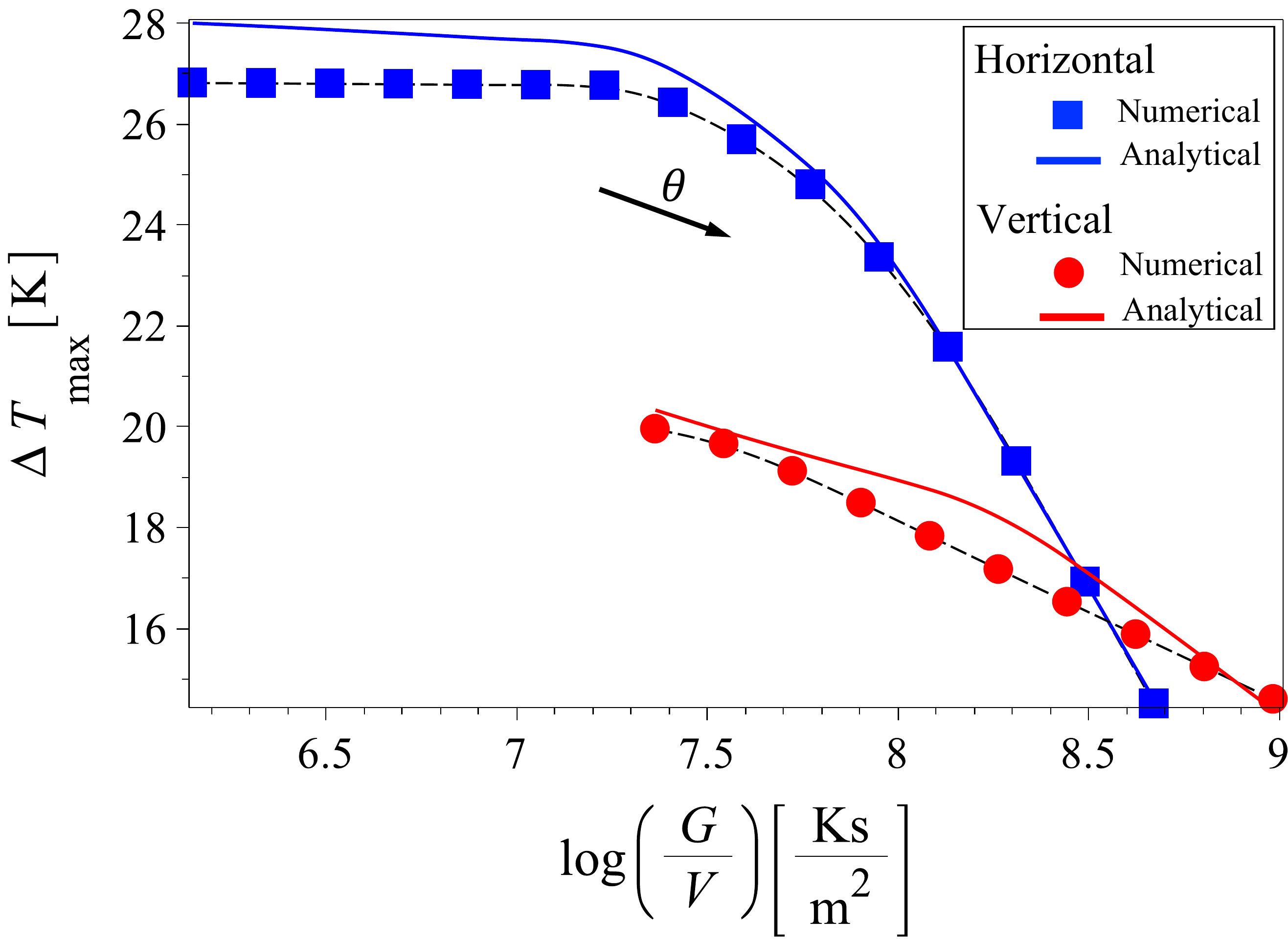}
\caption{Effect of solidification parameters on maximum constitutional undercooling for the horizontal and vertical build directions.  The corresponding analytical solutions for horizontal (\textcolor{blue}{\protect\solidrule}), and vertical (\textcolor{red}{\protect\solidrule}) are calculated for a range of ($G/V$) parameters extracted from the simulations.}
\label{fig:10}  
\end{figure}
The numerically calculated $\Delta T_{\rm max}$ for the range of $G$, $V$ parameters calculated along the solid-liquid interface for both build directions (Fig.~\ref{fig:9}) are plotted in Fig.~\ref{fig:10}. As a comparison, the corresponding analytical solutions (Eq.~(\ref{eq:DTmax})) for both build directions and for the same range of $G$, $V$ parameters are also presented in Fig.~\ref{fig:10}. The data shows that the analytical solutions are consistent with our numerically extracted results of maximum undercooling versus $G/V$. 

These results indicate that the maximum undercooling in both horizontal and vertical samples increases for a decreasing ratio of $G/V$, which in turn corresponds to decreasing of the angle $\theta$. This supports the hypothesis that there is a higher density of nucleated grains in the vicinity of the trailing edge of the melt pool, which decreases quickly towards the leading edge of the melt pool, as shown in Fig.~{\ref{fig:7}}(a-b)(\RNum{1}).   

It is noted that, the maximum undercooling in the horizontal case, varies over a broader range of $G/V$, and can achieve significantly larger values for smaller $G/V$ ratios compared to the vertical build direction.  By changing the building direction from vertical to horizontal, the normal to the solid-liquid interface becomes more tilted away from the building direction (i.e. smaller $\theta$ angles with respect to X-axis), which gives rise to larger $V$ and smaller $G$ values and thus lowers $G/V$, and results in larger $\Delta T_{\rm max}$.  The higher values of $\Delta T_{\rm max}$ in turn result in increasing the probability of equiaxed grain nucleation.  

%%%%%%%%%
\section{Conclusions}
\label{conc}
In this work, we conducted a numerical study to investigate the effect of building direction on the microstructure evolution of an LPBF-AlSi10Mg alloy. The thermal profiles of LPBF-AlSi10Mg for two horizontal and  vertical printing strategies were modelled by conducting finite element analysis heat transfer simulations. A multi-order parameter type phase field model was used to study grain growth under conditions analogous to LPBF for the dilute limit of AlSi10Mg, using underlying thermal conditions obtained from the heat transfer modelling for different build directions.  Our phase field model incorporated a stochastic noise-stimulated nucleation mechanism during free growth to simulate the spontaneous nuclei formation in the solidification process. 

The main conclusions of this study are summarised as follows:
\begin{itemize}
    \item The accuracy of our model to predict morphological transitions (such as columnar-to-equiaxed transition (CET)) was benchmarked by contrasting our numerical finding with predictions from a previously developed steady-state model for the CET. 
    
    \item Modelling the solidification behaviour of the single powder layer showed that grain growth for both build directions follows a similar pattern, where two downward (type A) and upward (type B) fronts of columnar grains merge at the trailing edge of the molten pool, shaping a zipper-like structure. The undercooled liquid between two approaching fronts then promotes further equiaxed nucleation events in these regions (type C). 
    
    \item  The formation of the zipper structure is the direct consequence of the thermal boundary and cooling conditions of  LPBF-AlSi10Mg that were dictated by the geometry of the deposited powder layer. 
  
    \item Numerical analysis of grain growth for different build directions indicate the strong effect of the melt pool morphology on constitutional undercooling, which in turns influences the solidification regime. 
    
    \item The maximum undercooling ($\Delta T_{\rm max}$) along the melt pool boundary was computed numerically in both horizontal and vertical samples. These were compared to an analytical prediction developed for this process, and found to be in reasonably good quantitative agreement and very good qualitative agreement.
    
    \item The analysis of $\Delta T_{\rm max}$ suggests that changing the build direction from vertical to horizontal increases the angle between maximal grain growth direction and building direction (i.e. Z-axis), thereby giving rise to higher constitutional undercooling, which in turn increases the probability of equiaxed grain nucleation.   
    
    \item Consistent with experimental observations, the number of equiaxed nucleation events we observe in our simulations is appreciably higher in the horizontal sample compared to the vertical sample, the former of which shows a higher equiaxed to columnar ratio.
\end{itemize}

The thermodynamics model of the alloy that we used in this study, although sufficient for qualitative study of the effect of build direction on microstructure evolution,  does not give a correct estimate of the local solute composition and the resulting dendritic tip undercooling for nucleation, and may fail to accurately predict morphological transitions (such as CET) in LPBF-AlSi10Mg. Therefore, a non-dilute binary phase field free energy must be developed for a more quantitative study.  

Furthermore, in order to better examine the dependence of CET on the building direction, we expect that it would be better to use the more quantitative CET model of Gaumann, Trivedi, Kurz (GTK) model.

Finally, the LPBF process for the range of processing parameters presented here falls in quasi-rapid solidification regime. Therefore, we expect that solute trapping due to high growth rates should also be incorporated into a more complete future phase-field study of this process in experimentally relevant applications \cite{pinomaa2019quantitative}.
%%%
\medskip

%%%
\section*{Acknowledgments}
The authors would like to thank Natural Sciences and Engineering Research Council of Canada (NSERC) grant number RGPIN-2016-04221, New Brunswick Innovation Foundation (NBIF) grant number RIF2017-071, Atlantic Canada Opportunities Agency (ACOA)- Atlantic Innovation Fund (AIF) project number 210414, and Mitacs Accelerate Program grant number IT10669 for providing sufficient funding to execute this work. The authors would also like to acknowledge the Canada Research Chairs Program for partially funding for this work. We also thank HPC McGill Centre (www.hpc.mcgill.ca), ACENET (www.ace-net.ca), and Compute Canada (www.computecanada.ca) for computing resources. \copyright Her Majesty the Queen in Right of Canada, as represented by the Minister of Natural Resources, 2019.\\
%%%

%%%
\bibliography{refrences}

%%%

%%%%%%%%%%%%%%%%%%%%%%%%%%%%%%%%%%%%%%%%%%%%%%%%%%%%%
\end{document}